\documentclass[aps,preprintnumbers,amsmath,amssymb,twocolumn,tightenlines,superscriptaddress]{revtex4}

\usepackage{graphicx}
\usepackage{epsfig}

\newcommand{\be}{\begin{eqnarray}}
\newcommand{\ee}{\end{eqnarray}}

 \newcommand{\gsim}{\mathrel{\hbox{\rlap{\lower.55ex \hbox {$\sim$}}
                   \kern-.3em \raise.4ex \hbox{$>$}}}}
\newcommand{\lsim}{\mathrel{\hbox{\rlap{\lower.55ex \hbox {$\sim$}}
                   \kern-.3em \raise.4ex \hbox{$<$}}}}

\newcommand{\ba}{\begin{eqnarray}}
\newcommand{\ea}{\end{eqnarray}}

\begin{document}


\title{Probing the Color Structure of the Perfect QCD Fluids \\ 
via Soft-Hard-Event-by-Event Azimuthal Correlations}
\author{Shuzhe Shi} 
\address{Physics Department and Center for Exploration of Energy and Matter,
Indiana University, 2401 N Milo B. Sampson Lane, Bloomington, IN 47408, USA.}
\author {Jinfeng Liao} 
\address{Physics Department and Center for Exploration of Energy and Matter,
Indiana University, 2401 N Milo B. Sampson Lane, Bloomington, IN 47408, USA.}
\author {Miklos Gyulassy}  
\address{Nuclear Science Division, Lawrence Berkeley National Laboratory, Berkeley, CA 94720, USA.}
\address{Pupin Lab MS-5202, Department of Physics, Columbia University, New York, NY 10027, USA.}
\address{Institute of Particle Physics and Key Laboratory of Quark \& Lepton Physics (MOE), Central China Normal University, Wuhan, 430079, China.}

\begin{abstract}
  We develop a comprehensive dynamical framework, CIBJET, to calculate
  on an event-by-event basis the dependence of correlations between
  soft ($p_T<2$ GeV) and hard ($p_T> 10$ GeV) azimuthal flow angle 
  harmonics on the color composition of near-perfect QCD fluids
  produced in high energy nuclear collisions at RHIC and LHC. CIBJET
  combines consistently predictions of event-by-event VISHNU2+1
  viscous hydrodynamic fluid fields with CUJET3.1
  predictions of event-by-event jet quenching. We
  find that recent correlation data favor a temperature dependent color
  composition including
  bleached chromo-electric $q(T)+g(T)$ components and an
  emergent chromo-magnetic
  degrees of freedom $m(T)$ consistent with non-perturbative lattice QCD
  information  in the confinement/deconfinement temperature range.

\end{abstract}

\maketitle


{\it Introduction and Conclusion.} At extremely high temperature $T\sim 10^{12}$ K a new form of strongly interacting QCD matter referred to as a
strongly coupled  quark-gluon plasma (sQGP) is created. The sQGP was the primordial form of matter occupying the early Universe micro-seconds after the Big Bang, and is now recreated in heavy ion collision experiments at the Relativistic Heavy Ion Collider (RHIC) and the Large Hadron Collider (LHC)~\cite{Gyulassy:2004zy,Shuryak:2004cy,Jacobs:2004qv,Muller:2012zq,Shuryak:2014zxa}. Lattice QCD\cite{Bazavov:2009zn,Borsanyi:2010bp} predicts a rapidly varying composition of the microscopic color degrees of freedom in the critical crossover confinement/deconfinement temperature range $100<T\sim T_c <300$ MeV.
The experimental discovery that the sQGP produced at RHIC and LHC  exhibits signatures of near
perfect fluidity with shear viscosity to entropy density ratio 
close to the quantum bound $\eta/s\sim 0.1-0.2$
\cite{Danielewicz:1984ww,Kovtun:2004de} offers a unique opportunity to test lattice
QCD predictions of the $T\sim T_c$  color composition suggested by 
numerical simulations of  the Polyakov loop value, quark number susceptibilities, chromo-electric and magnetic screening masses, and the QCD equation of state. In particular,
the jet quenching transport coefficient, $\hat{q}(T,E)$, that controls
the quenching pattern of high energy jets and their hadronic fragments
is sensitive to the microscopic color composition of the produced QCD fluids
as we demonstrated previously at least in
simplified event-averaged fluid geometries\cite{Xu:2014ica,Xu:2014tda,Xu:2015bbz,Shi:2017onf}. We recently completed a comprehensive 
global
$\chi^2$ analysis with CIBJET of all soft-hard azimuthal correlation data  
on the jet fragment nuclear modification factor, $R_{AA}(p_T,\phi;\sqrt{s},b)$,
in the center-of-mass energy range  $\sqrt{s}=0.2-5.02$ ATeV and impact parameter range $b=1\sim 10$ fm that will be reported in depth elsewhere\cite{Shi:2018}. This Letter highlights the major conclusions of that detailed analysis.

The high transverse momentum chromo-electric quark and gluon jets 
are produced before the soft fluid and suffer energy loss as well as transverse
diffusion due to jet-medium interactions along its path through
the evolving inhomogeneous and expanding
fluid medium.  Any  nontrivial temperature dependence of the color
composition between color electric and color magnetic charges
can be expected to leave an imprint in the azimuthal
 quenching pattern as emphasized in \cite{Liao:2008dk,Betz:2012qq,Betz:2014cza,Zhang:2012ie,Zhang:2012ha,Li:2014hja,Das:2015ana}. 
  The pattern of energy loss in transverse momentum $p_T$ and azimuthal angle $\phi$
  provides tomographic information~\cite{Gyulassy:2000gk} via the harmonic decomposition of
  the nuclear modification factor: 
\begin{eqnarray}
 R_{AA} (p_T; \phi) &=& 
  R_{AA}(p_T) {\big [}1  + 2v_2(p_T)\cos(2\phi-2\Psi_2) \nonumber \\
&&  \quad + 2v_3(p_T)\cos(3\phi-3\Psi_3) + ... {\big ]}
\end{eqnarray}
where the coefficient $v_n$ in the angular distribution  of the jet fragments is essentially measured with respect to azimuthal 
flow harmonics $v_n^{soft} e^{i \Psi_n}$ of the soft $(p_T<2~{\rm GeV})$ hadronic fragments from the QCD fluid.  As emphasized in \cite{Noronha-Hostler:2016eow,Betz:2016ayq} 
it is essential to utilize the soft-hard correlated experimental definition of
the jet harmonic coefficients 
$v_n(p_T)\equiv\langle v_n^{soft}v_n^{hard}(p_T)\rangle/\sqrt{\langle (v_n^{soft})^2\rangle}$ where 
$\langle\cdots\rangle$ refers to the event-by-event ensemble average in a particular centrality class.

\begin{figure}  
\begin{center}
\includegraphics[width=0.36\textwidth,height=0.33\textwidth]{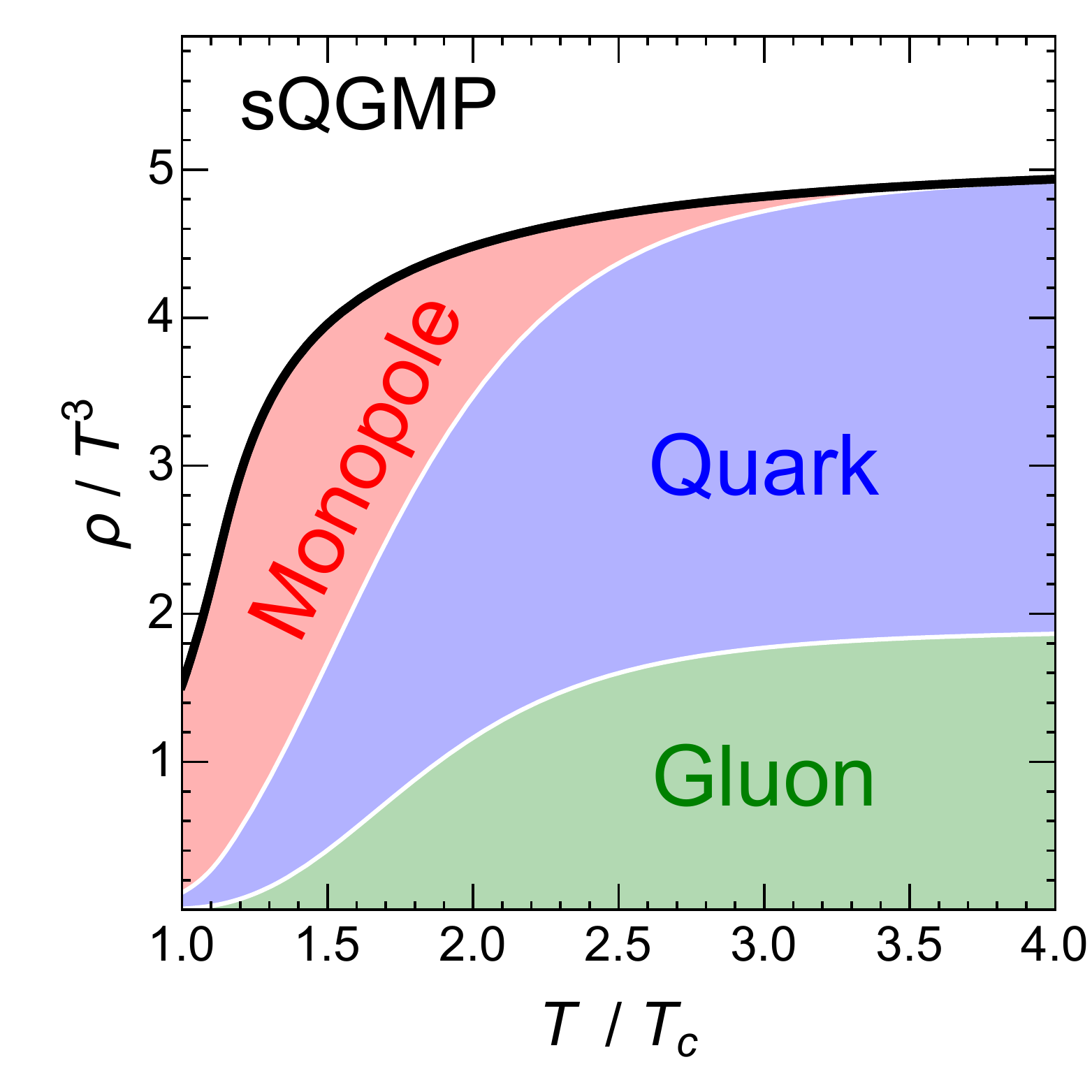}
\includegraphics[width=0.36\textwidth,height=0.33\textwidth]{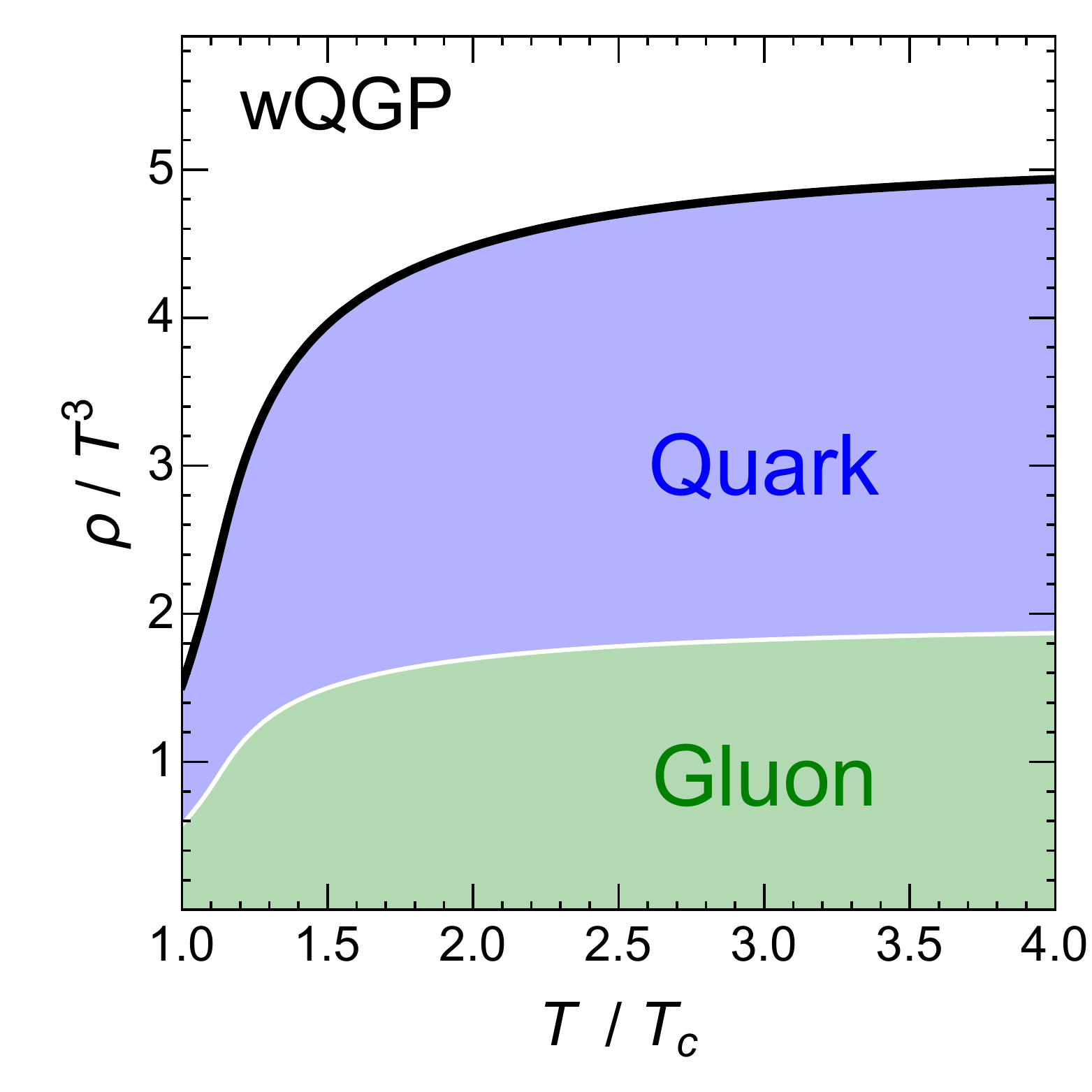}
\vspace{-0.1in}
\caption{(color online) The two color composition models: semi-Quark-Gluon-Monopole Plasma (sQGMP) versus the perturbative quark+gluon plasma (wQGP). In each plot, the solid black curve is from the lattice data for entropy density normalized by cubic temperature, representing the total constituent densities in the plasma. At a given temperature, the vertical width of different color bands reflects the density fraction of each component (quark, gluon or monopole). Compared with wQGP, the sQGMP features a strong suppression of chromo-electric densities accompanied by a rapid increase of  monopole density when approaching $T_c$ from above.}
\vspace{-0.3in}
\label{fig0}
\end{center}
\end{figure}

Event-by-event fluctuations strongly influence soft flow
observables. Odd harmonics are entirely determined by fluctuations.
The transfer of the pattern of soft azimuthal flow fluctuations
onto the pattern of hard azimuthal fluctuations requires a combined
simultaneous quantitative description of both soft long wavelength and
hard short wavelength dynamics. In \cite{Noronha-Hostler:2016eow,Betz:2016ayq} the first successful simultaneous account of Soft-Hard observables
$R_{AA},v_2(p_T)$ and $v_3(p_T)$ was demonstrated using the
ebe-vUSPHydro+BBMG framework. In that framework a parametric BBMG  energy loss model with linear path-length dependence was used
that could however not be further exploited to constrain
the color composition of the QCD fluid. The CIBJET model solves that problem
by combining iEBE-VISHNU+CUJET3.1 models and is the first combined soft+hard framework with sufficient generality to test different color composition
models. The default composition option, referred to as a semi-Quark-Gluon-Monopole Plasma (sQGMP), involves suppressed color electric quark and gluon
degrees of freedom $q(T)+g(T)$ as well as the emergent color magnetic monopole degrees of freedom, $m(T)$,
with their temperature dependence as implied  by available lattice QCD data. This is in contrast to the perturbative QCD/HTL composition, referred as wQGP, 
that is limited to only color screened electric $q(T)+g(T)$
quark and gluon quasi-parton degrees of freedom. The two color composition models are illustrated in Fig.~\ref{fig0}.

In \cite{Xu:2014tda,Xu:2015bbz,Shi:2017onf} we showed that the sQGMP composition accounts well
for RHIC and LHC data at least in the simplified approximation
when event-averaged smooth geometries are assumed. In this Letter we
show that the CIBJET event-by-event generalization of our previous work
does not change our central conclusion that the sQGMP color composition is
preferred over the perturbative QCD/HTL composition
that is limited to only color screened electric $q(T)+g(T)$
quark and gluon quasi-parton degrees of freedom
that are also not consistent with lattice QCD data in the critical crossover temperature range. In this Letter we further show
that another composition model~\cite{Zakharov:2014bca},  referred to as mQGP, that includes magnetic monopoles based on lattice estimations on top of quarks and gluons but  does not
suppress $ q(T)+g(T)$,  is also inconsistent with the $v_2$ data once the coupling is adjusted to reproduce $R_{AA}$. We conclude this Letter by showing that
the sQGMP jet transport coefficient $\hat{q}(T,E)$ peaks near $T_c$ with sufficient strength as to provide a natural  dynamical explanation of how the QCD fluid $\eta/s\approx T^3/\hat{q}$ could approach the perfect fluid bound near $T_c$
due to the emergent $m(T)$ component.

{\it The CIBJET framework.---} 
The bulk evolution in CIBJET is simulated on an event-by-event basis, by using the  viscous hydrodynamic simulation code VISHNU~\cite{Shen:2014vra} which has been widely used and well vetted at both RHIC and the LHC.  Two types of initial conditions will be employed, the Monte-Carlo Glauber (with a corresponding $\eta/s=0.1$ for the hydro) or the Trento (with $\eta/s=0.2$ and $p=0$)~\cite{Moreland:2014oya}, both of which are phenomenologically viable for describing soft bulk observables.

The high $p_T$ jet energy loss component of CIBJET is based on the CUJET3.1 model developed over the past several years~\cite{Buzzatti:2011vt,Xu:2014ica,Xu:2014tda,Xu:2015bbz,Shi:2017onf}. It is a jet energy loss simulation framework built upon a non-perturbative microscopic model for the hot partonic medium as a semi-quark-gluon-monopole plasma (sQGMP). The energy loss calculation includes both the Thoma-Gyulassy  elastic energy loss~\cite{Thoma:1990fm} for collisional processes as well as the dynamical DGLV opacity expansion theory~\cite{Gyulassy:2000er,Djordjevic:2003zk,Wicks:2005gt} for radiative processes. The most nontrivial aspect of the CUJET3.1 is the chromo structure of the QGP medium when approaching $T_c\sim \Lambda_{QCD}$,  which integrates two key features arising from nonperturbative dynamics pertaining to the confinement transition. The first is the suppression of chromo-electric degrees of freedom from high T toward $T_c$, as proposed and studied in the so-called semi-QGP model~\cite{Hidaka:2008dr,Hidaka:2009ma,Lin:2013efa}. The second is the emergence of the chromo-magnetic degrees of freedom, i.e. the magnetic monopoles, which become dominant in the near-$T_c$ regime and eventually reach condensation to enforce confinement at $T<T_c$, known as the ``magnetic scenario'' and studied extensively~\cite{Liao:2006ry,Liao:2007mj,Liao:2008jg,Ratti:2008jz,DAlessandro:2007lae,Bonati:2013bga}.  Detail descriptions of the CUJET3 component can be found in e.g.~\cite{Xu:2014tda,Xu:2015bbz,Shi:2017onf,Shi:2018}.

There are two key parameters in this framework: $\alpha_c$ which is the nonperturbative coupling strength at the transition temperature scale $T_c\simeq 160~\rm MeV$; and $c_m$ which controls the magnetic screening mass and sensitively influences the scattering rates  involving the magnetic component.  A recent  comprehensive comparison of  the model calculations (based on smooth-hydro background) for $R_{AA}$ and $v_2$ with extensive  data from RHIC to LHC~\cite{Acharya:2018qsh,ATLAS:2017rmz,Khachatryan:2016odn,Sirunyan:2017pan,Adam:2016izf},  has allowed us to optimize these parameters. The $\alpha_c$ most sensitively controls overall opaqueness and is fixed as $\alpha_c=0.9$; the $c_m$ strongly influences anisotropy $v_2$  and is fixed as $c_m=0.25$ for Glauber geometry while $c_m=0.22$ for Trento geometry~\cite{Shi:2017onf,Shi:2018}.

{\it A Unified Soft-Hard Description with CIBJET.---} With the above CIBJET setup, we've performed the highly demanding event-by-event simulations for sophisticated jet energy loss calculations at the LHC energy. It may be noted that computation for each centrality costs about sixty-thousand cpu hours. In each event, the bulk medium evolves from the hydro component while on top of that about $5\times10^5$ jet paths are sampled for energy loss calculation (as well as accounting for path fluctuations and gluon emission sampling). With such computing power, we are able to quantitatively explore both soft and hard observables in a unified simulation framework and to answer the aforementioned pressing questions.

\begin{figure}[!htbt]
\begin{center}
\includegraphics[width=0.48\textwidth]{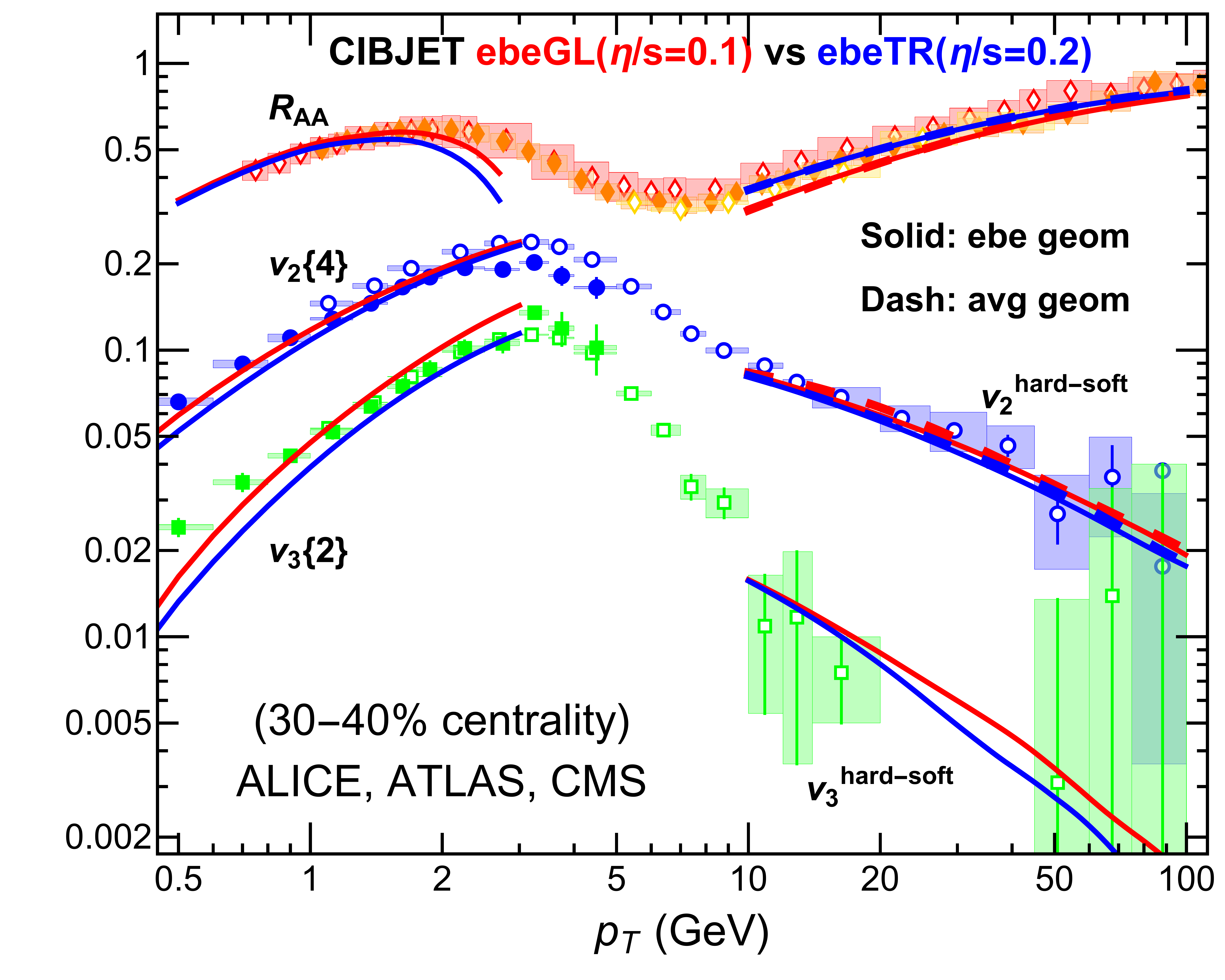} 
\vspace{-0.1in}
\caption{(color online) The nuclear modification factor $R_{AA}$ as well as the second and third harmonic coefficients $v_2$ \& $v_3$ of the final hadron azimuthal distribution as functions of $p_T$ for 30-40\% Pb+Pb collisions at 5.02~ATeV. The solid curves are from event-by-event calculations while the dashed from  average-geometry. The CIBJET results in both soft and hard regions, with either Monte-Carlo Glauber (red)  or Trento (blue) initial conditions, are in excellent agreement with experimental data from  ALICE, ATLAS and CMS~\cite{Acharya:2018qsh,ATLAS:2017rmz,Khachatryan:2016odn,Sirunyan:2017pan,Adam:2016izf}.}
\vspace{-0.2in}
\label{fig1}
\end{center}
\end{figure}

The CIBJET results for nuclear modification factor $R_{AA}$ as well as the second and third harmonic coefficients $v_2$ \& $v_3$ of the final hadron azimuthal distribution are shown in Fig.\ref{fig1}  for 30-40\% Pb+Pb collisions at 5.02~ATeV. 
It should be particularly noted that the anisotropy observables are computed in the same way as the experimental analysis on an event-wise basis.  
The solid curves are from event-by-event CIBJET  with either Monte-Carlo Glauber (red) initial condition and $\eta/s=0.1$   or Trento (blue) initial condition and $\eta/s=0.2$, while the dashed curves are single-shot calculations with the corresponding averaged smooth geometry. Observables in the soft region ($p_T\lsim 2~\rm GeV$) are computed from the hydrodynamic component while observables in the hard region ($p_T\gsim 10~\rm GeV$) are computed from the jet energy loss component. The CIBJET results in both soft and hard regions, spanning a broad transverse momentum window from $0.5~\rm GeV$ to $100~\rm GeV$, are in excellent agreement with available experimental data from  ALICE, ATLAS and CMS collaborations at the LHC~\cite{Acharya:2018qsh,ATLAS:2017rmz,Khachatryan:2016odn,Sirunyan:2017pan,Adam:2016izf}. The results for the $v_3$ at high $p_T$ deserve special note, which could not possibly be computed without  event-by-event simulations and which serves as a further independent test of the CIBJET's phenomenological success. 

\begin{figure}[!htbt]
\begin{center}
\includegraphics[width=0.45\textwidth]{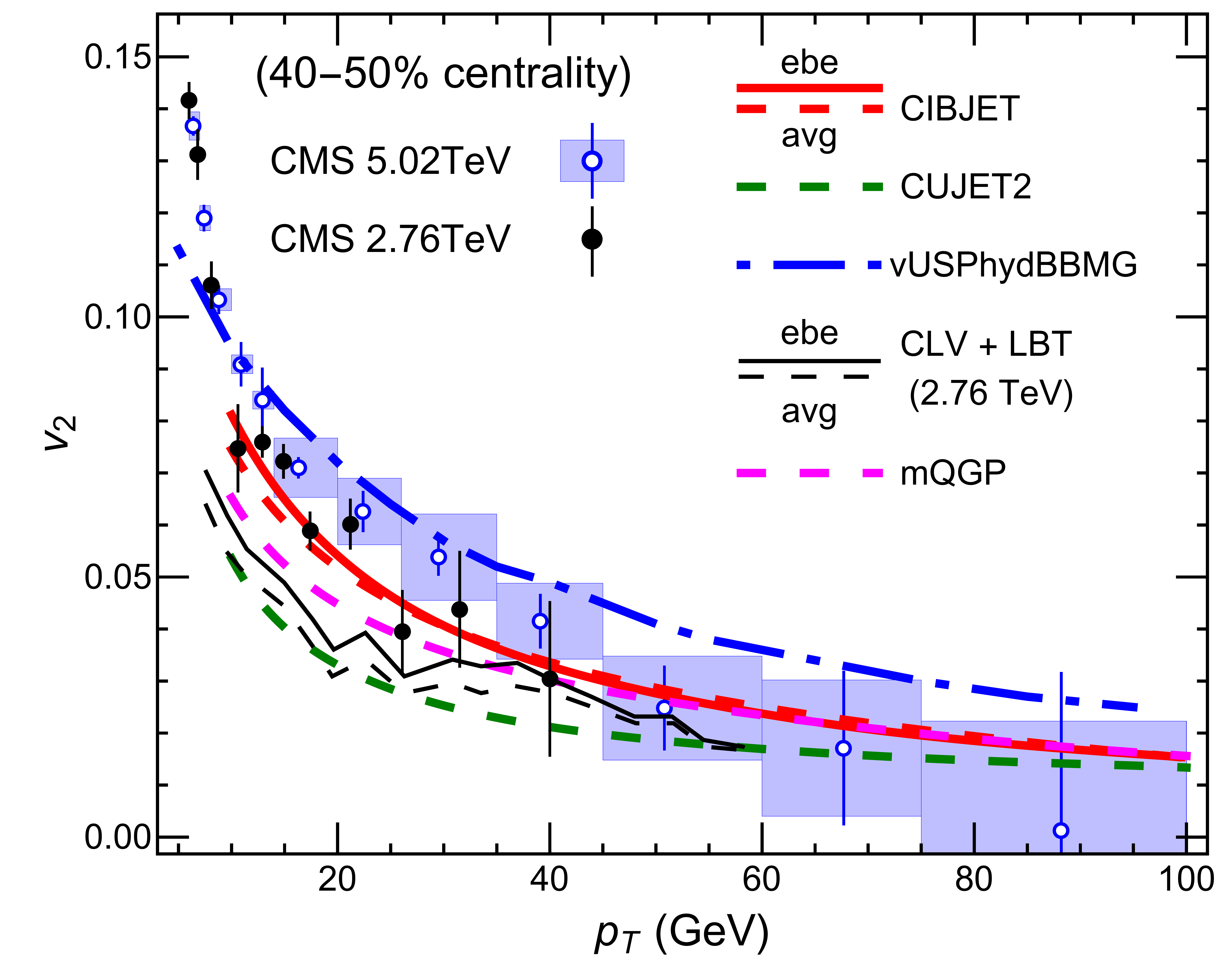} 
\vspace{-0.1in}
\caption{(color online) A comparison of  $v_2$ at high $p_T$  from different models with CMS data~\cite{Chatrchyan:2012xq,Sirunyan:2017pan} for 40-45\% Pb+Pb collisions at 2.76~ATeV and 5.02~ATeV, including: CIBJET (red) with event-by-event (solid) or average geometry (dashed), CLV+LBT (black) with event-by-event (solid) or average geometry (dashed), CUJET2 (dashed green), vUSPhydBBMG (dash-dotted blue) and mQGP (dashed magenta).} 
\vspace{-0.2in}
\label{fig2}
\end{center}
\end{figure}

One important issue is whether the high $p_T$ anisotropy $v_2$ from event-by-event calculations could indeed be strongly enhanced from that obtained with average smooth geometry in the same model. A hint for such enhancement was recently reported  from the vUSPhydBBMG model~\cite{Noronha-Hostler:2016eow}, which simulates jet energy loss based upon simple parameterized polynomial dependence on path-length, medium temperature and parton energy on top of an event-by-event hydro background. From CIBJET results in Fig.~\ref{fig1}, however, no significant difference has been detected between the event-by-event case and the average geometry case for either Glauber or Trento initial conditions.   
To further investigate this issue, let us focus on $v_2$ at high $p_T$  and compare a number of models in Fig.~\ref{fig2}. In addition to the CIBJET and vUSPhydBBMG models, three more models are included for this comparison: (1) the CUJET2 model which has a similar DGLV framework as CIBJET but is based on a perturbative quark-gluon medium with HTL resummation~\cite{Xu:2014ica}; (2) the CLV+LBT model which uses a higher-twist-formalism-based linearized Boltzmann approach in a perturbative quark-gluon medium with simulations on top of the CLV viscous hydro background~\cite{Cao:2017umt,Cao:2017hhk}; (3) the Zakharov's mQGP model~\cite{Zakharov:2014bca}, which computes the energy loss in the BDMPS-Z formalism based on a medium that adds  magnetic monopoles on top of the usual perturbative quark-gluon sector. In between the CUJET2 or CLV+LBT and the CIBJET or mQGP, the main difference is that the latter two models' medium includes a chromo magnetic component. In between CIBJET and mQGP, the main difference is that the CIBJET has its chromo electric component being gradually suppressed toward lower temperature, while the mQGP has no suppression of the quark/gluon sector and directly adds an estimated monopole density~\cite{DAlessandro:2007lae,Bonati:2013bga}. All models  nicely describe the  same $R_{AA}$ data and their results for $v_2$ would provide the critical test. We note in passing other different  models not included in this comparison~\cite{Bianchi:2017wpt,Chien:2015vja,Djordjevic:2014tka,Rapp:2018qla}.

As seen from Fig.~\ref{fig2}, despite their significant difference in many aspects,  both CIBJET (red) and CLV+LBT (black) models demonstrate very small difference between their respective average-geometry results (dashed curves) and event-by-event (solid curves) results. This observation indicates at a limited role of event-by-event fluctuations in the quantitative evaluation of high $p_T$ anisotropy $v_2$. In comparison with CMS data, the CLV+LBT (black) and the CUJET2 (green dashed) models, both based a perturbative medium of QGP with HTL resummation, under-predict the $v_2$ values. The CIBJET (red) and Zakharov mQGP (magenta) models, both including a strong magnetic component near $T_c$ and thus enhancing late time energy loss, give much larger $v_2$ than the CUJET2 or CLV+LBT model, with the CIBJET in good agreement with data. 
This comparative study clearly demonstrates the differentiating power of the high $p_T$ anisotropy observable, and strongly suggests two important points: (1) the event-by-event fluctuations have limited impact on the hard sector $v_2$ values; (2) the inclusion of chromo magnetic component for the medium enhances the hard sector $v_2$ and is crucial for describing experimental data~\cite{Liao:2008dk,Zhang:2012ie,Xu:2014tda,Ramamurti:2017zjn}. \\

 {\it  Transport properties and color structure of QGP.---} Further insights on the viability of a unified and consistent understanding of the soft and hard sectors together can be obtained by  investigating the corresponding soft and hard transport properties of QGP. That is, one could try to calculate the $\hat{q}_F/T^3$ and $\eta/s$ for a given QGP medium model whose parameters have been calibrated with data. Here we explore three models with distinct chromo structure: a wQGP medium (as in CUJET2) with only chromo-electric component, a mQGP medium with unsuppressed chromo-electric component plus an added magnetic component, as well as a sQGMP medium (as in CIBJET) with suppressed chromo-electric component and an emergent magnetic component. These coefficients are computed  by properly synthesizing  the contributions from all components to the momentum-square transfer with a jet (in the case of $\hat{q}_F/T^3$) or to the scattering cross-sections (in the case of $\eta/s$). The detailed formulae can be found in e.g. \cite{Xu:2014tda,Xu:2015bbz}.

\begin{figure*} [hbt!]
\begin{center} 
\includegraphics[width=0.4\textwidth]{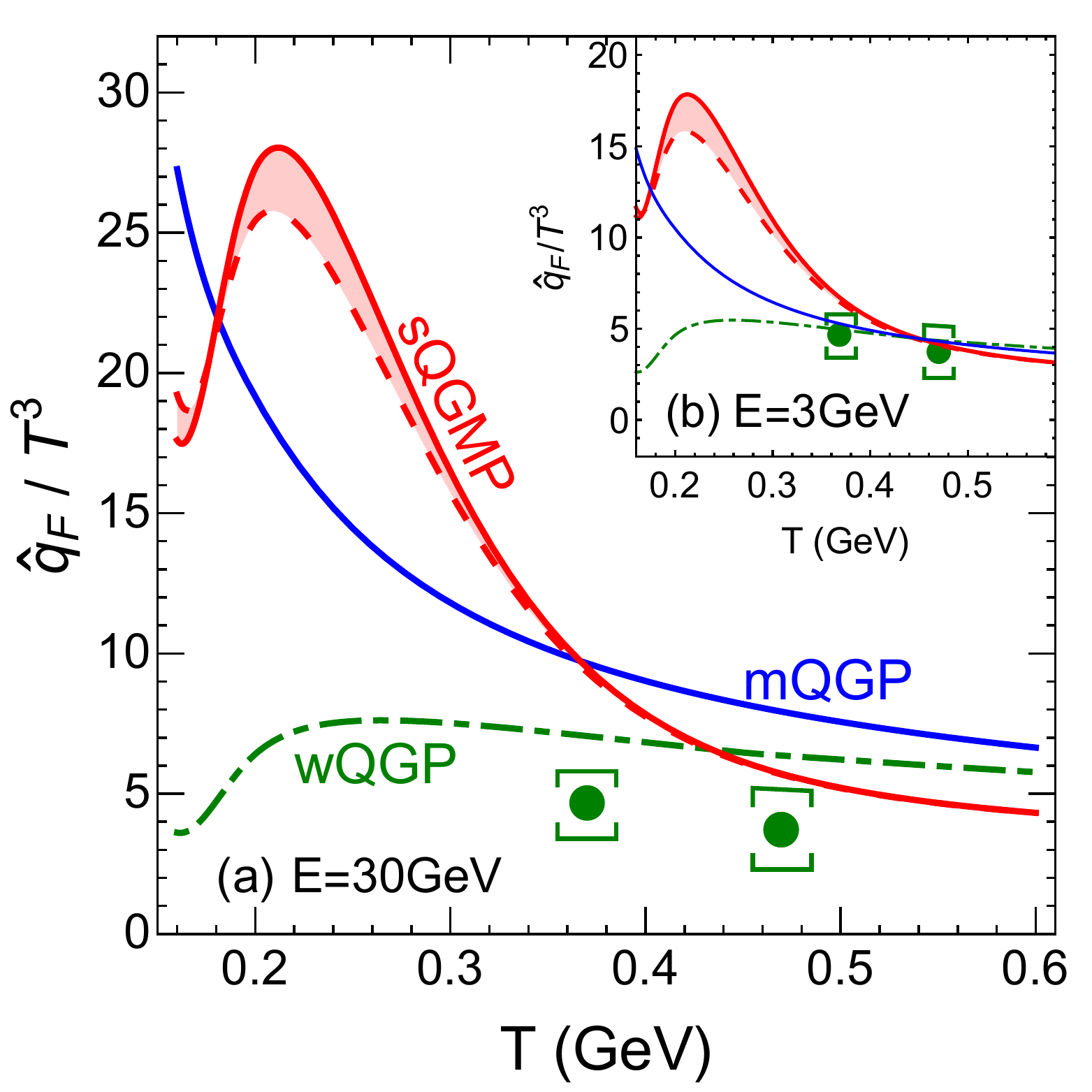} \hspace{0.1in}
\includegraphics[width=0.4\textwidth]{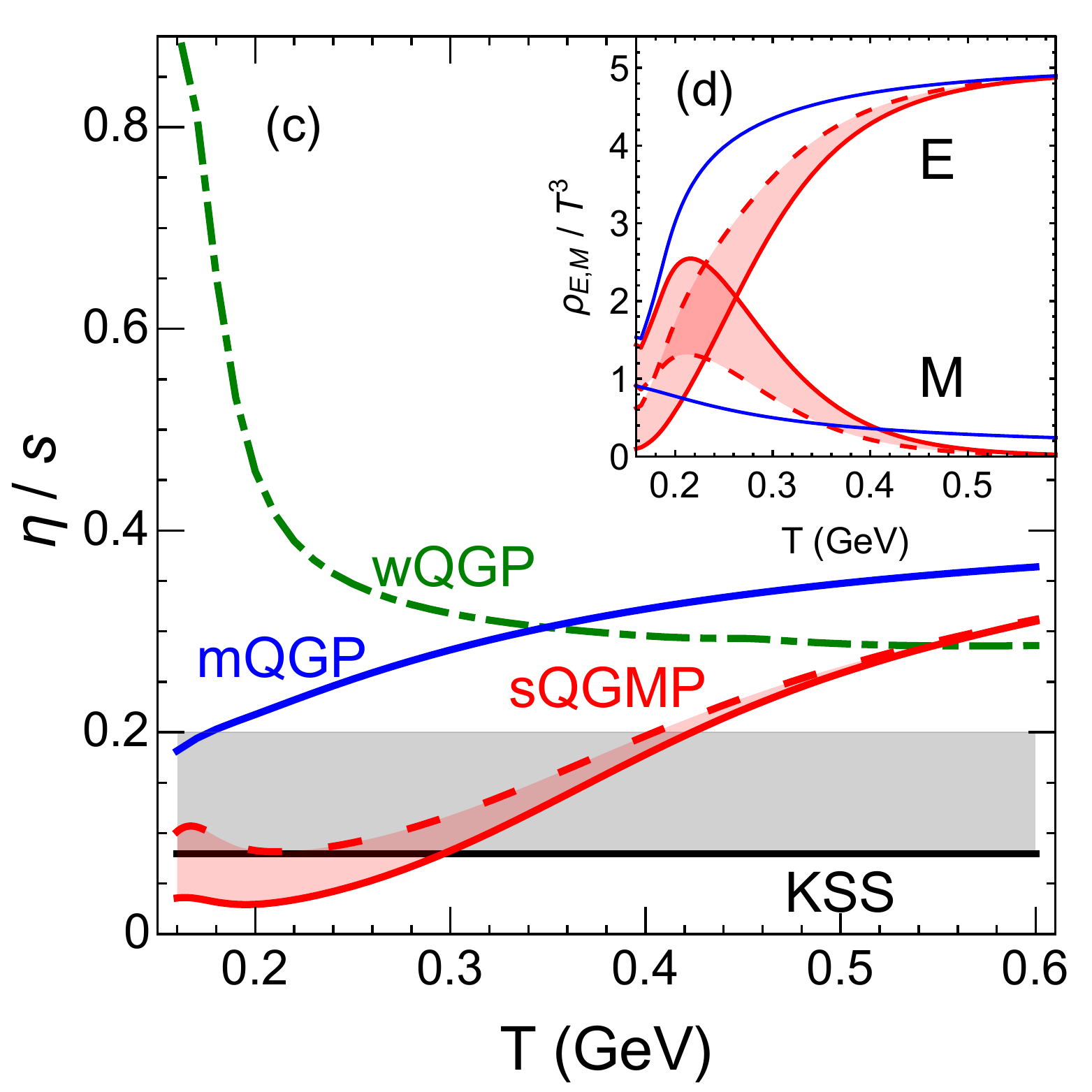}
\caption{(color online)
(left) The jet transport coefficient $\frac{\hat{q}_F}{T^3}$  for jet energy (a) $E=30~\rm GeV$ or (b) $E=3~\rm GeV$; (right) the shear viscosity coefficient $\frac{\eta}{s}$ in (c) for models with their different color electric and magnetic density decompositions shown in (d).  The models include the wQGP model (dash-dotted green), the mQGP model (solid blue) as well as the sQGMP model (red) with either $\chi_T^L$ scheme (solid) or $\chi_T^u$ scheme (dashed). The two green dots in (a)(b) are values from JET collaboration extraction for jet energy $E=10~\rm GeV$~\cite{Burke:2013yra}. The KSS black line indicates $\eta/s=1/(4\pi)$ holographic bound while the grey band indicates the average $\eta/s$ value range constraints from bulk hydro phenomenology.  (See text for details.)  
} \label{fig3} \vspace{-0.2in}
\end{center}
\end{figure*}  

The results are shown in Fig.~\ref{fig3}. Both sQGMP and mQGP models show a strong enhancement  of $\hat{q}_F/T^3$ and the decrease of $\eta/s$ in the near $T_c$ regime, an important feature that is absent in the wQGP and is due to the emergence of the magnetic component. Such nontrivial near-$T_c$ features are much stronger in sQGMP than in mQGP, which could be understood from the fact that the magnetic component is more dominant in sQGMP (i.e. the ratio of  magnetic density to electric density is larger, see panel (d) of Fig.~\ref{fig3}). In terms of hard sector, it is already apparent from Fig.~\ref{fig2} that the $v_2$ at high $p_T$  favors the sQGMP model. For the soft sector,  the $\eta/s$ comparison also clearly favors  the sQGMP model, leading to  $\eta/s\sim (0.1\sim 0.2)$ in the relevant temperature regime which are precisely the needed values (for either Glauber or Trento initial conditions) for hydro calculations to correctly produce the bulk soft anisotropy observables $v_2$ and $v_3$ in Fig.~\ref{fig1}. We note in passing that  the sQGMP transport coefficients around $T_c$ are close to the values suggested from strongly coupled field theories via AdS/CFT approach~\cite{Kovtun:2004de,Liu:2006ug}.

Within the sQGMP scenario, we've further studied two slightly different suppression scheme for the chromo-electric component. The $\chi_T^L$ scheme uses the lattice-computed Polyakov loop as a ``penalty'' for color charge to characterize the suppression of quark sector  as in the original semi-QGP. The $\chi_T^u$ scheme instead uses the lattice-computed quark number susceptibilities to quantify the suppression of quark sector. In both schemes the suppression of gluon sector is based on Polyakov loop as in semi-QGP. The main difference  is that there is stronger (faster) suppression in the $\chi_T^L$ scheme than the $\chi_T^u$ scheme of chromo-electric component from high T toward low T: see the solid versus dashed curves in the panel (d) of Fig.~\ref{fig3}.  With both schemes phenomenologically viable, 
 the $\chi_T^u$ scheme seems preferred by virtue of consistency with the KSS bound $\eta/s\ge 1/4\pi$~\cite{Kovtun:2004de}.

{\it Conclusion.---} In summary, we've established a comprehensive CIBJET framework as a sophisticated and realistic event-by-event simulation tool that allows for a unified, quantitative and consistent description of both soft and hard sector observables $\left( R_{AA}\otimes v_2 \otimes v_3 \right)$ across a wide span of transverse momentum from $\sim 0.5~\rm GeV$ to $\sim 100~\rm GeV$ and in excellent agreement with experimental data. Such phenomenological success strongly suggests at a highly nontrivial color structure of the near-perfect QCD fluid as a semi-Quark-Gluon-Monopole Plasma (sQGMP), which is in line with the variation of color degrees of freedom as suggested by lattice QCD for the temperature regime most relevant to current heavy ion collision experiments.  Remarkably, the sQGMP  also provides a dynamical explanation of the  temperature-dependent jet transport coefficient $\frac{\hat{q}_F}{T^3}$  and shear viscosity coefficient $\frac{\eta}{s}$ that are mutually consistent as well as consistent with extracted values from phenomenology. 

 \vspace{0.1in}

 {\bf Acknowledgements.---}The authors thank B. Betz, S. Cao, U. Heinz P. Jacobs, A. Majumder, J. Noronha, J. Noronha-Hostler and X.-N. Wang for useful discussions and communications. The authors are particularly grateful to Jiechen Xu and Alessandro Buzzatti for their crucial contributions to the development of CUJET framework. This work is supported  by the National Science Foundation under Grant No. PHY-1352368.


\vspace{-0.2in}

 \begin{thebibliography}{99}


\bibitem{Gyulassy:2004zy} 
  M.~Gyulassy and L.~McLerran,
  Nucl.\ Phys.\ A {\bf 750}, 30 (2005). 
  
\bibitem{Shuryak:2004cy} 
  E.~V.~Shuryak,
  Nucl.\ Phys.\ A {\bf 750}, 64 (2005).


\bibitem{Jacobs:2004qv} 
  P.~Jacobs and X.~N.~Wang,
  Prog.\ Part.\ Nucl.\ Phys.\  {\bf 54}, 443 (2005)
  doi:10.1016/j.ppnp.2004.09.001
  [hep-ph/0405125].
  
\bibitem{Muller:2012zq} 
  B.~Muller, J.~Schukraft and B.~Wyslouch,
  Ann.\ Rev.\ Nucl.\ Part.\ Sci.\  {\bf 62}, 361 (2012). 
  
\bibitem{Shuryak:2014zxa} 
  E.~Shuryak,
  Rev.\ Mod.\ Phys.\  {\bf 89}, 035001 (2017). 
  
\bibitem{Bazavov:2009zn} 
  A.~Bazavov {\it et al.},
  ``Equation of state and QCD transition at finite temperature,''
  Phys.\ Rev.\ D {\bf 80}, 014504 (2009)
  doi:10.1103/PhysRevD.80.014504
  [arXiv:0903.4379 [hep-lat]].
  
\bibitem{Borsanyi:2010bp} 
  S.~Borsanyi {\it et al.} [Wuppertal-Budapest Collaboration],
  JHEP {\bf 1009}, 073 (2010)
  doi:10.1007/JHEP09(2010)073
  [arXiv:1005.3508 [hep-lat]].
  
\bibitem{Danielewicz:1984ww} 
  P.~Danielewicz and M.~Gyulassy,
  Phys.\ Rev.\ D {\bf 31}, 53 (1985).
  
\bibitem{Kovtun:2004de} 
  P.~Kovtun, D.~T.~Son and A.~O.~Starinets,
  Phys.\ Rev.\ Lett.\  {\bf 94}, 111601 (2005)
  doi:10.1103/PhysRevLett.94.111601
  [hep-th/0405231].
  
\bibitem{Xu:2014ica} 
  J.~Xu, A.~Buzzatti and M.~Gyulassy,
  JHEP {\bf 1408}, 063 (2014)
  doi:10.1007/JHEP08(2014)063
  [arXiv:1402.2956 [hep-ph]].


\bibitem{Xu:2014tda} 
  J.~Xu, J.~Liao and M.~Gyulassy,
  Chin.\ Phys.\ Lett.\  {\bf 32}, no. 9, 092501 (2015)
  doi:10.1088/0256-307X/32/9/092501
  [arXiv:1411.3673 [hep-ph]].
  
\bibitem{Xu:2015bbz} 
  J.~Xu, J.~Liao and M.~Gyulassy,
  JHEP {\bf 1602}, 169 (2016)
  doi:10.1007/JHEP02(2016)169
  [arXiv:1508.00552 [hep-ph]].
  
\bibitem{Shi:2017onf} 
  S.~Shi, J.~Xu, J.~Liao and M.~Gyulassy,
  Nucl.\ Phys.\ A {\bf 967}, 648 (2017)
  doi:10.1016/j.nuclphysa.2017.06.037
  [arXiv:1704.04577 [hep-ph]].
  
  \bibitem{Shi:2018} 
   S.~Shi, J.~Liao and M.~Gyulassy, to appear soon. 
     
\bibitem{Liao:2008dk} 
  J.~Liao and E.~Shuryak,
  Phys.\ Rev.\ Lett.\  {\bf 102}, 202302 (2009)
  doi:10.1103/PhysRevLett.102.202302
  [arXiv:0810.4116 [nucl-th]].
  
     
\bibitem{Betz:2012qq} 
  B.~Betz and M.~Gyulassy,
  Phys.\ Rev.\ C {\bf 86}, 024903 (2012)
  doi:10.1103/PhysRevC.86.024903
  [arXiv:1201.0281 [nucl-th]].  
  
  
\bibitem{Betz:2014cza} 
  B.~Betz and M.~Gyulassy,
  JHEP {\bf 1408}, 090 (2014)
  Erratum: [JHEP {\bf 1410}, 043 (2014)]
  doi:10.1007/JHEP10(2014)043, 10.1007/JHEP08(2014)090
  [arXiv:1404.6378 [hep-ph]].


\bibitem{Zhang:2012ie} 
  X.~Zhang and J.~Liao,
  Phys.\ Rev.\ C {\bf 89}, no. 1, 014907 (2014)
  doi:10.1103/PhysRevC.89.014907
  [arXiv:1208.6361 [nucl-th]].
  
  
\bibitem{Zhang:2012ha} 
  X.~Zhang and J.~Liao,
  Phys.\ Rev.\ C {\bf 87}, 044910 (2013)
  doi:10.1103/PhysRevC.87.044910
  [arXiv:1210.1245 [nucl-th]].

\bibitem{Li:2014hja} 
  D.~Li, J.~Liao and M.~Huang,
  Phys.\ Rev.\ D {\bf 89}, no. 12, 126006 (2014)
  doi:10.1103/PhysRevD.89.126006
  [arXiv:1401.2035 [hep-ph]].
  
    
\bibitem{Das:2015ana} 
  S.~K.~Das, F.~Scardina, S.~Plumari and V.~Greco,
  Phys.\ Lett.\ B {\bf 747}, 260 (2015). 
    
\bibitem{Gyulassy:2000gk} 
  M.~Gyulassy, I.~Vitev and X.~N.~Wang,
  Phys.\ Rev.\ Lett.\  {\bf 86}, 2537 (2001)
  doi:10.1103/PhysRevLett.86.2537
  [nucl-th/0012092].


\bibitem{Noronha-Hostler:2016eow} 
  J.~Noronha-Hostler, B.~Betz, J.~Noronha and M.~Gyulassy,
  Phys.\ Rev.\ Lett.\  {\bf 116}, no. 25, 252301 (2016)
  doi:10.1103/PhysRevLett.116.252301
  [arXiv:1602.03788 [nucl-th]].
  
\bibitem{Betz:2016ayq} 
  B.~Betz, M.~Gyulassy, M.~Luzum, J.~Noronha, J.~Noronha-Hostler, I.~Portillo and C.~Ratti,
  Phys.\ Rev.\ C {\bf 95}, no. 4, 044901 (2017)
  doi:10.1103/PhysRevC.95.044901
  [arXiv:1609.05171 [nucl-th]].
  
  
\bibitem{Zakharov:2014bca} 
  B.~G.~Zakharov,
  JETP Lett.\  {\bf 101}, no. 9, 587 (2015)
  [Pisma Zh.\ Eksp.\ Teor.\ Fiz.\  {\bf 101}, no. 9, 659 (2015)]
  doi:10.1134/S0021364015090131
  [arXiv:1412.6287 [hep-ph]].
  

\bibitem{Shen:2014vra} 
  C.~Shen, Z.~Qiu, H.~Song, J.~Bernhard, S.~Bass and U.~Heinz,
  Comput.\ Phys.\ Commun.\  {\bf 199}, 61 (2016)
  doi:10.1016/j.cpc.2015.08.039
  [arXiv:1409.8164 [nucl-th]].

\bibitem{Moreland:2014oya} 
  J.~S.~Moreland, J.~E.~Bernhard and S.~A.~Bass,
  Phys.\ Rev.\ C {\bf 92}, no. 1, 011901 (2015)
  doi:10.1103/PhysRevC.92.011901
  [arXiv:1412.4708 [nucl-th]].


\bibitem{Buzzatti:2011vt} 
  A.~Buzzatti and M.~Gyulassy,
  Phys.\ Rev.\ Lett.\  {\bf 108}, 022301 (2012)
  doi:10.1103/PhysRevLett.108.022301
  [arXiv:1106.3061 [hep-ph]].
  
 
  
\bibitem{Thoma:1990fm} 
  M.~H.~Thoma and M.~Gyulassy,
  Nucl.\ Phys.\ B {\bf 351}, 491 (1991).

  
  
\bibitem{Gyulassy:2000er} 
  M.~Gyulassy, P.~Levai and I.~Vitev,
  Nucl.\ Phys.\ B {\bf 594}, 371 (2001). 
  
\bibitem{Djordjevic:2003zk} 
  M.~Djordjevic and M.~Gyulassy,
  Nucl.\ Phys.\ A {\bf 733}, 265 (2004). 
 
\bibitem{Wicks:2005gt} 
  S.~Wicks, W.~Horowitz, M.~Djordjevic and M.~Gyulassy,
  Nucl.\ Phys.\ A {\bf 784}, 426 (2007). 
  

\bibitem{Hidaka:2008dr} 
  Y.~Hidaka and R.~D.~Pisarski,
  Phys.\ Rev.\ D {\bf 78}, 071501 (2008)
  doi:10.1103/PhysRevD.78.071501
  [arXiv:0803.0453 [hep-ph]].
  
\bibitem{Hidaka:2009ma} 
  Y.~Hidaka and R.~D.~Pisarski,
  Phys.\ Rev.\ D {\bf 81}, 076002 (2010)
  doi:10.1103/PhysRevD.81.076002
  [arXiv:0912.0940 [hep-ph]].
  
\bibitem{Lin:2013efa} 
  S.~Lin, R.~D.~Pisarski and V.~V.~Skokov,
  Phys.\ Lett.\ B {\bf 730}, 236 (2014)
  doi:10.1016/j.physletb.2014.01.043
  [arXiv:1312.3340 [hep-ph]].
 
  
\bibitem{Liao:2006ry} 
  J.~Liao and E.~Shuryak,
  Phys.\ Rev.\ C {\bf 75}, 054907 (2007)
  doi:10.1103/PhysRevC.75.054907
  [hep-ph/0611131].
  
\bibitem{Liao:2007mj} 
  J.~Liao and E.~Shuryak,
  Phys.\ Rev.\ C {\bf 77}, 064905 (2008)
  doi:10.1103/PhysRevC.77.064905
  [arXiv:0706.4465 [hep-ph]].
  
\bibitem{Liao:2008jg} 
  J.~Liao and E.~Shuryak,
  Phys.\ Rev.\ Lett.\  {\bf 101}, 162302 (2008)
  doi:10.1103/PhysRevLett.101.162302
  [arXiv:0804.0255 [hep-ph]].
  
  
\bibitem{Ratti:2008jz} 
  C.~Ratti and E.~Shuryak,
  Phys.\ Rev.\ D {\bf 80}, 034004 (2009)
  doi:10.1103/PhysRevD.80.034004
  [arXiv:0811.4174 [hep-ph]].
  
\bibitem{DAlessandro:2007lae} 
  A.~D'Alessandro and M.~D'Elia,
  Nucl.\ Phys.\ B {\bf 799}, 241 (2008)
  doi:10.1016/j.nuclphysb.2008.03.002
  [arXiv:0711.1266 [hep-lat]].
  
\bibitem{Bonati:2013bga} 
  C.~Bonati and M.~D'Elia,
  Nucl.\ Phys.\ B {\bf 877}, 233 (2013)
  doi:10.1016/j.nuclphysb.2013.10.004
  [arXiv:1308.0302 [hep-lat]].
 
 
\bibitem{Acharya:2018qsh} 
  S.~Acharya {\it et al.} [ALICE Collaboration],
  arXiv:1802.09145 [nucl-ex].
  
\bibitem{ATLAS:2017rmz} 
  The ATLAS collaboration [ATLAS Collaboration],
  ATLAS-CONF-2017-012.
  
\bibitem{Khachatryan:2016odn} 
  V.~Khachatryan {\it et al.} [CMS Collaboration],
  JHEP {\bf 1704}, 039 (2017)
  doi:10.1007/JHEP04(2017)039
  [arXiv:1611.01664 [nucl-ex]].
  
\bibitem{Sirunyan:2017pan} 
  A.~M.~Sirunyan {\it et al.} [CMS Collaboration],
  Phys.\ Lett.\ B {\bf 776}, 195 (2018)
  doi:10.1016/j.physletb.2017.11.041
  [arXiv:1702.00630 [hep-ex]].
  
\bibitem{Adam:2016izf} 
  J.~Adam {\it et al.} [ALICE Collaboration],
  Phys.\ Rev.\ Lett.\  {\bf 116}, no. 13, 132302 (2016)
  doi:10.1103/PhysRevLett.116.132302
  [arXiv:1602.01119 [nucl-ex]].
   
   
\bibitem{Cao:2017umt} 
  S.~Cao, L.~G.~Pang, T.~Luo, Y.~He, G.~Y.~Qin and X.~N.~Wang,
  Nucl.\ Part.\ Phys.\ Proc.\  {\bf 289-290}, 217 (2017).

\bibitem{Cao:2017hhk} 
  S.~Cao, T.~Luo, G.~Y.~Qin and X.~N.~Wang,
  Phys.\ Lett.\ B {\bf 777}, 255 (2018).

\bibitem{Bianchi:2017wpt} 
  E.~Bianchi, J.~Elledge, A.~Kumar, A.~Majumder, G.~Y.~Qin and C.~Shen,
  arXiv:1702.00481 [nucl-th].


\bibitem{Chien:2015vja} 
  Y.~T.~Chien, A.~Emerman, Z.~B.~Kang, G.~Ovanesyan and I.~Vitev,
  Phys.\ Rev.\ D {\bf 93}, no. 7, 074030 (2016). 
  
\bibitem{Djordjevic:2014tka} 
  M.~Djordjevic, M.~Djordjevic and B.~Blagojevic,
  Phys.\ Lett.\ B {\bf 737}, 298 (2014). 
  
\bibitem{Rapp:2018qla} 
  R.~Rapp {\it et al.},
  arXiv:1803.03824 [nucl-th].

\bibitem{Chatrchyan:2012xq} 
  S.~Chatrchyan {\it et al.} [CMS Collaboration],
  Phys.\ Rev.\ Lett.\  {\bf 109}, 022301 (2012)
  doi:10.1103/PhysRevLett.109.022301
  [arXiv:1204.1850 [nucl-ex]].
  
  
\bibitem{Ramamurti:2017zjn} 
  A.~Ramamurti and E.~Shuryak,
  Phys.\ Rev.\ D {\bf 97}, no. 1, 016010 (2018)
  doi:10.1103/PhysRevD.97.016010
  [arXiv:1708.04254 [hep-ph]].
  
\bibitem{Burke:2013yra} 
  K.~M.~Burke {\it et al.} [JET Collaboration],
  Phys.\ Rev.\ C {\bf 90}, no. 1, 014909 (2014)
  doi:10.1103/PhysRevC.90.014909
  [arXiv:1312.5003 [nucl-th]].
  
  
  
\bibitem{Liu:2006ug} 
  H.~Liu, K.~Rajagopal and U.~A.~Wiedemann,
  Phys.\ Rev.\ Lett.\  {\bf 97}, 182301 (2006)
  doi:10.1103/PhysRevLett.97.182301
  [hep-ph/0605178].
   
   
\end{thebibliography}
\end{document}